\documentclass[12pt]{article}
\pdfoutput=1
\usepackage{verbatim}
\usepackage{amsmath}
\usepackage{amssymb}
\usepackage{amsthm}
\usepackage{slashed}
\usepackage{amsfonts}
\usepackage[utf8]{inputenc}
\usepackage[normalem]{ulem}
\usepackage{tikz}
\usepackage{dsfont}
\usepackage{braket}
\usepackage{youngtab}
\usepackage{amsfonts}
\usepackage{mathrsfs}
\usepackage{array}
\usepackage{float}
\usepackage{graphicx, rotating}
\usepackage{epstopdf}
\usepackage{epsfig}
\usepackage{cite}
\usepackage{latexsym}
\usepackage{color}
\usepackage{bm}
\usepackage{slashed}
\usepackage[pagebackref=true]{hyperref}
\definecolor{rossos}{cmyk}{0,1,1,0.55}
\definecolor{bluscuro}{rgb}{0.15, 0.2, .85}
\definecolor{bluchiaro}{cmyk}{1,.3,0.,0.1}
\hypersetup{colorlinks, citecolor=bluscuro, linkcolor=bluscuro, urlcolor=bluscuro}

\def\be{\begin{equation}}
\def\ee{\end{equation}}
\def\ba#1\ea{\begin{align}#1\end{align}}
\def\bg#1\eg{\begin{gather}#1\end{gather}}
\def\bm#1\em{\begin{multline}#1\end{multline}}
\def\bmd#1\emd{\begin{multlined}#1\end{multlined}}

\def\p{\partial}

\def\({\left(}
\def\){\right)}
\def\[{\left[}
\def\]{\right]}
\def\<{\langle}
\def\>{\rangle}

\newcommand{\dd}{\text{d}}

\newcommand{\bfig}{\begin{figure}\begin{center}}
\newcommand{\efig}{\end{center}\end{figure}}
\newcommand{\bi}{\begin{itemize}}
\newcommand{\ei}{\end{itemize}}

\newcommand{\pt}{\partial}

\theoremstyle{definition}

\begin{document}
%
%
\begin{titlepage}
	\begin{flushright}
		IFT-UAM/CSIC-20-166
	\end{flushright}
	\vspace{.3in}

		\vspace{1cm}
	\begin{center}
		{\Large\bf\color{black} A Generalized Momentum/Complexity Correspondence}\\
		
		\bigskip\color{black}
		\vspace{1cm}{
			{\large J.~L.~F. Barb\'on$^a$, J. Mart\'{\i}n-Garc\'{\i}a$^b$  and M.~Sasieta$^a$}
			\vspace{0.3cm}
		} \\[7mm]
		$^a$	{\it {Instituto de F\'{i}sica Te\'orica IFT-UAM/CSIC,\\ c/ Nicol\'as Cabrera 13, Universidad Aut\'onoma de Madrid, 28049, Madrid, Spain}}\\[10pt]
		$^b$	{\it {Instituto Galego de F\'isica de Altas Enerx\'ias (IGFAE), \\
					Universidade de Santiago de Compostela,
				E-15782, Santiago de Compostela, Spain}}\\[10pt]
		
		{\it E-mail:} \href{mailto:jose.barbon@csic.es}{\nolinkurl{jose.barbon@csic.es}}, \href{mailto:javier.martingarcia1@gmail.com}{\nolinkurl{javier.martingarcia1@gmail.com}},  \href{mailto:martin.sasieta@csic.es}{\nolinkurl{martin.sasieta@csic.es}}
	\end{center}
	
	\bigskip

\vspace{1cm}

\begin{abstract}

Holographic complexity, in the guise of the Complexity = Volume prescription, comes equipped with a natural correspondence between its rate of growth 
and the average infall momentum of matter in the bulk. This Momentum/Complexity correspondence can be related to an integrated version of the momentum constraint of general relativity.   In this paper we propose a generalization, using the full Codazzi equations as a starting point, which successfully accounts for purely gravitational contributions to infall momentum. The proposed formula is explicitly checked in an exact pp-wave solution of the vacuum Einstein equations. 

\end{abstract}
\end{titlepage}

\tableofcontents

\vspace{1cm}

\section{Introduction}
\label{sec:intro}

A recurrent idea since the early days of holography is a somewhat implicit relation between   notions of `complexity' of quantum states and the degree  of  `gravitational clumpling'  in the bulk  \cite{Susskindhologram, Verlinde}. Recently, these intuitions came gradually into focus in low-dimensional models \cite{SusskindFall,Susskindchargedfall,SusskindNewton,Magan, MaldacenaLin, Mousatov} and other AdS/CFT examples  \cite{ushell, prd, SusskindZhaoMomentum,VaydiaI,VaydiaII}, leading to momentum/complexity relations of the generic form
\be\label{gen}
{\dd {\cal C} \over \dd t } \sim P_{\rm infall}
\;,
\ee
equating  the rate of complexity growth with a suitable `infall' momentum measuring the rate of gravitational clumping of matter in the bulk. The claimed ability to see `deep in the bulk' is one of the prime motivations for the introduction of quantum complexity in general discussions of holography. Therefore,  a correspondence of the type \eqref{gen} offers an interesting quantitative handle on this fundamental property, since the infall momentum flux is locally computable in the bulk. 

Equations of the form \eqref{gen} have been known to appear in various analyses of gravitational collapse based on spherical shells (cf. \cite{ushell, Susskindswitchbacks, VaydiaI, VaydiaII}). It was suggested in \cite{ushell} that a fairly general form of the correspondence should exist, beyond the specific dynamical details of the various shell models. This expectation was born out  by the results in  \cite{prd}, showing that an equation of the form \eqref{gen} is always present, as a consequence of the momentum constraint of general relativity (GR), provided  the complexity is defined with the Complexity=Volume prescription \cite{SusskindEntnotEnough,Susskindlectures,StanfordShockWave,Susskindswitchbacks,Robertslocalized}. More precisely, if ${\cal C}$ is defined as an extremal codimension-one volume, one finds 
\be\label{pvc} 
{\dd {\cal C} \over \dd t} =   P_C [\Sigma_t] + R_C [\Sigma_t] 
\;,
\ee
where  $\Sigma_t$ is the extremal surface used to compute the volume complexity at a given time $t$. The infall momentum $P_C [\Sigma_t]$ is defined in terms of the matter energy-momentum $T_{\mu\nu}$  by the expression 
\be\label{pc}
P_C [\Sigma_t] = -\int_{\Sigma_t} N^\mu \;T_{\mu\nu}\; C^\nu \;.
\ee
Here 
$N^\mu$ denotes the future-directed  unit normal  to $\Sigma_t$ and $C^\mu$
is a vector field, tangent to $\Sigma_t$, which defines what is precisely meant by the `infall' momentum component. Asymptotically, the field $C$  must be chosen to be radial, inward pointing and with a modulus given by the radius of the angular sphere at long distances. 
A certain degree of arbitrariness in the choice of $C$ in the bulk is reflected in the existence of the  `remainder', 
\be\label{reminder}
R_C [\Sigma_t] = -{1\over 8\pi G} \int_{\Sigma_t} K^{ab} \,\nabla_a \,C_b \;,
\ee
where $K^{ab}$ denotes the extrinsic curvature of $\Sigma_t$. 
The reminder term  vanishes whenever the $C$-field can be chosen as a conformal Killing vector field, something that happens in any $2+1$ dimensional bulk or any spherically symmetric state in arbitrary dimensions. When the state is such that $R_C [\Sigma_t] =0$, we say that the momentum/complexity correspondence (PVC) is exact. 

The physical interpretation of the infall momentum in \eqref{pc} is greatly clarified by considering the Newtonian limit, another situation where the remainder can be neglected.  We can model  the matter as  a set of Newtonian particles moving in a fixed background metric, which can be taken to be the vacuum AdS manifold with curvature radius $\ell$. For particles localized in a region  of size much smaller than the AdS curvature radius, we can approximate the background metric as flat Minkowski spacetime, for which the $C$-field becomes purely radial, with component $C^r = - r/\ell$ in standard polar coordinates. Under these conditions \eqref{pc} simplifies to 
\be\label{pcn}
{\dot {\cal C}}_{\rm Newtonian} \approx P_{\rm infall} =  -{1\over \ell} \sum_i {\bf x}_i \cdot {\bf p}_i \;, 
\ee
that is to say, the infall momentum is a sort of radial analog of the angular momentum. 
More emphatically, we can define a `degree of clumping' of the particle system as a kind of `spherical moment of inertia', 
\be\label{clump}
{\cal I}_{\rm clump} = -{1\over 2\ell} \sum_i m_i \,{\bf x}_i^2 \;,
\ee 
and conclude that the `Newtonian complexity' is determined by such degree of clumping, up to a constant: 
\be\label{pcc}
{\cal C}_{\rm Newtonian} = {\cal C}_0 + {\cal I}_{\rm clump} \;.
\ee
Thus, this relation  formalizes the intuitive notion that `matter clumping' increases complexity. 

Notwithstanding these important cases that show an exact or approximate PVC correspondence, the remainder term in \eqref{pvc} does not vanish in general. The simplest and more important instance  occurs when considering pure gravity solutions. Gravitational wave scattering with black hole formation is the crucial example of a process with a non-vanishing rate of complexity growth, which must come entirely from the remainder term. The purpose of this paper is to propose  a generalization of \eqref{pvc} in which the notion of  `infall momentum'  is suitably generalized, in a way that can handle pure gravity solutions. The very existence of such a generalization is quite remarkable, given that no strictly local definition of energy-momentum exists for the purely gravitational degrees of freedom. 

This paper is organized as follows. In section \ref{sec:GPVC} we review the PVC correspondence introduced in \cite{prd} and explain its proposed generalization. In section \ref{sec:PPWaves} we test our proposal by an explicit calculation in the case of an exact gravitational wave background. In section \ref{sec:conclusions} we state our conclusions and offer various avenues for future research.

\section{Generalized PVC Correspondence}
\label{sec:GPVC}

\begin{figure}[h]
	\centering
	\includegraphics[width = .3\textwidth]{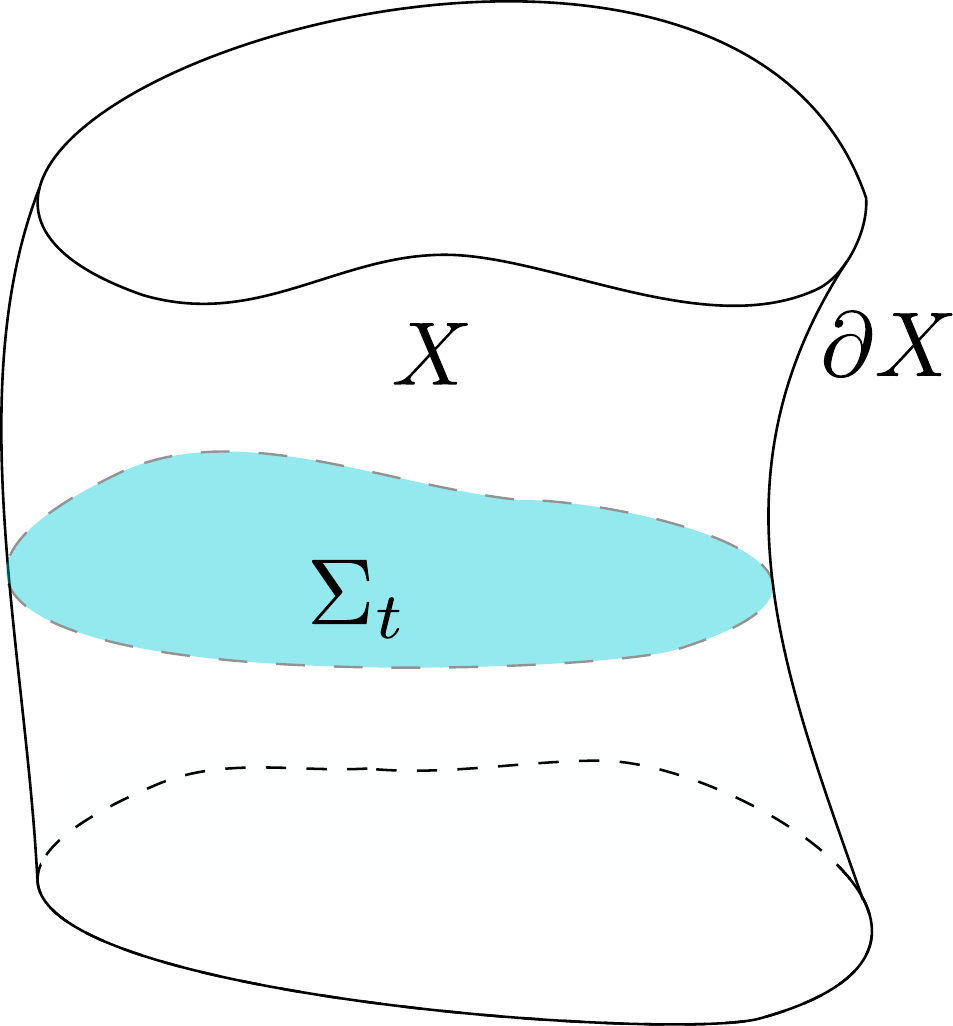}
	\caption{\textit{Schematic representation of the region $X$, whose spatial boundary $\pt X$ has cylindrical topology.}}
	\label{fig:cylinder}
\end{figure}

In order to illustrate the general structure of PVC correspondences, we start by introducing some notation. Let the metric $g_{\mu\nu}$ be a solution of Einstein equations with  energy-momentum tensor $T_{\mu\nu}$.    Given a portion of spacetime $X$  which fills a cylinder $\pt X$  in $d+1$ dimensions, as in Figure \ref{fig:cylinder}, we shall define its  volume-complexity  by the following procedure.
First, we consider a Hamiltonian foliation of $\pt X$  by spacelike surfaces of spherical topology,  denoted $\pt \Sigma_t$, and labeled by a time variable $t$. Associated to each $\pt\Sigma_t$ we pick a smooth extension into the bulk, $\Sigma_t$, defined as a maximal-volume hypersurface with boundary $\pt \Sigma_t$. 

We shall refer to the intrinsic and extrinsic geometry of $\Sigma_t$ as the `bulk state' at time $t$. In the classical approximation for the bulk dynamics, this amounts to the specification of both  the
induced metric on $\Sigma_t$, denoted $h_{ab}$, and the extrinsic curvature $K_{ab}$ of its embedding into $X$. These quantities have standard definitions in terms of the embedding fields $X^\mu (y^a)$, the associated frames $e^\mu_a = \pt_a X^\mu$,  and the unit, future-directed normal  $N^\mu$. 
\be\label{defsi}
h_{ab} = g_{\mu\nu} \, e^\mu_a \,e^\nu_b  \;, \qquad K_{ab} =e^\mu_a \,e^\nu_b\; \nabla_\mu N_\nu  \;.
\ee
Greek indices refer to an arbitrary coordinate system $x^\mu$  of $X$ and are raised and lowered with $g_{\mu\nu}$, whereas Latin indices refer to the world-volume coordinates $y^a$ and are raised and lowered with $h_{ab}$. 

With these ingredients at hand, we can define the volume-complexity  at time $t$ as proportional to the volume of the extremal surface $\Sigma_t$: 
\be\label{vcc}
{\cal C} [\Sigma_t] = {d-1 \over 8\pi G\, b} \;{\rm Vol} [\Sigma_t].
\ee
In this expression, $b$ stands for an arbitrary length scale which is part of the normalization ambiguity of the complexity and represents an effective `box size' for the physics of interest. For instance, in AdS/CFT applications, one often sets $b$ to  the asymptotic AdS curvature radius, after \eqref{vcc} is conveniently renormalized at infinity. Alternatively,  for studies of black holes in flat space, setting $b$ to the scale of the event horizon proves more convenient (cf. \cite{VCnormalization}). 

We regard  the so-defined complexity as a measure of `quantum  complexity' in a hypothetical holographic description with geometrical data on $\pt X$, even if we do not know in general how to characterize such a theory. Strictly speaking, only when $\pt X$ is regarded as the conformal boundary of an asymptotically AdS spacetime, one has a natural candidate for the holographic theory living on $\pt X$ (in this case a CFT with certain dynamical requirements).  Even in this case, the relation of this `volume complexity' to other notions of complexity,  such as circuit or tensor network complexity, is tentative. On the other hand, codimension-one extremal surfaces with fixed boundary data are certain to define interesting quantities in the holographic dual which are at least analogous to computational complexity. The discussion in this paper takes place in bulk variables,  so that it is independent of the ultimate boundary interpretation of `volume complexity'. At any rate, it is natural to assume that these notions will continue to be relevant in any putative holographic description of general regions $X$. 

Since the hypersurfaces $\Sigma_t$ are defined as extremal,  the volume 
$$
{\rm Vol} \,[\Sigma] = \int_{\Sigma} \,\dd ^d y\,  \sqrt{{\rm det}(h_{ab})}
$$
is stationary at $\Sigma_t$, with respect to variations of the embedding functions, $\delta X^\mu$, with fixed Dirichlet conditons on $\partial \Sigma_t$. Conversely, if the embedding variation is free on $\pt X$, the first-order variation of the volume picks a boundary term of the form
\be\label{bt}
\delta V [\Sigma_t] = \int_{\pt \Sigma_t}  \dd S^a \,e_a^\mu \,\delta X_\mu \;. 
\ee
Considering now boundary variations that only shift the time variable, $\delta X^\mu = \delta t \,(\pt_t)^\mu$, where $\pt_t$ denotes the future-directed, time displacement vector, we obtain an equation for the rate of complexity growth which only depends on a boundary integral: 
\be\label{bint}
{\dd {\cal C}\over \dd t} = {d-1 \over 8\pi G \,b} \int_{\pt{ \Sigma_t}} (\pt_t)_{\Sigma_t} = {d-1 \over 8\pi G \,b} \int_{\pt{ \Sigma_t}} \dd S^a \,(\pt_t)_a \;. 
\ee
This expression  involves the pullback onto $\Sigma_t$ of the time displacement vector, with components  $(\pt_t)_a = e^\mu_a (\pt_t)_\mu$. Since the extremal surfaces $\Sigma_t$ are in one-to-one correspondence with the time variable $t$, we will often suppress the time label and consider the extremal surfaces $\Sigma$ as equivalent specifications of the time slice.  

We would like to emphasize that equation \eqref{bint} is valid for any  foliation of a smooth manifold $X$ by hypersurfaces anchored on $\partial X$. In particular, we do not require that $\pt X$  be a conformal  AdS  boundary.

\subsection{Restricted PVC from the Momentum Constraint of GR } 

The expression of  the complexity rate \eqref{bint} as a purely boundary integral suggests a general strategy to derive momentum/complexity correspondences. 
Given any `current' $J^a$ defined on $\Sigma$, which has the same boundary integral as \eqref{bint}, we can use Stokes' theorem to write the complexity rate  as a bulk integral of its `source' over the extremal surface:
\be\label{bu}
{\dd{\cal C}\over \dd t} = {d-1\over 8\pi G \,b} \int_{\pt{\Sigma}} \dd S_a\,J^a = 
{d-1\over 8\pi G \, b} \int_{{\Sigma}} \nabla_a J^a\;.
\ee
A judicious choice of $J^a$ would relate $\nabla^a J_a $ to a momentum density on $\Sigma$, and a natural candidate 
presents itself. Since the extremal surfaces are valid initial-value surfaces in the Hamiltonian formalism of GR, we can use the momentum constraint to guess the right form of the current $J^a$: given any Cauchy surface $\Sigma$, initial data $h_{ab}$ and $K_{ab}$ are constrained by the equation (cf. \cite{Poisson})
\be\label{mc}
\nabla^a K_{ab} - \nabla_b K = -8\pi G \,{\cal P}_b\;,
\ee
where  $K= h^{ab} K_{ab}$ and
\be\label{tmunu}
{\cal P}_b = -N^\mu\,T_{\mu\nu} \, e^\nu_b 
\ee
is the pulled-back momentum density through $\Sigma$. If $\Sigma$ is a maximal-volume hypersurface, then the extrinsic curvature is traceless, $K=0$, further simplifying the equation. 
Notice that $e^\nu$ and $N^\mu$ are orthogonal vectors, so that  \eqref{tmunu} is insensitive to any `dark energy' components of $T_{\mu\nu}$, proportional to $g_{\mu\nu}$. In the discussion that follows, any explicit occurrence of $T_{\mu\nu}$ will involve the combination  ${\cal P}_b$ so that, by convention,  we will talk about the `matter energy-momentum' in reference to localized degrees of freedom that can undergo gravitational clumping. 

\begin{figure}[h]
	\centering
	\includegraphics[width = .4\textwidth]{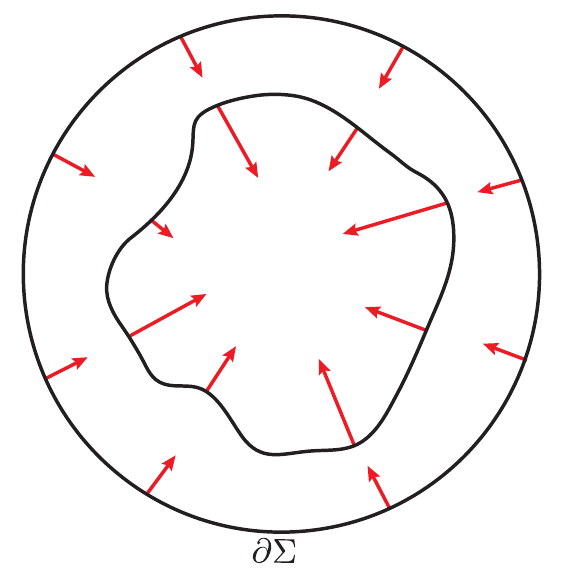}
	\caption{\textit{The canonical $C$-field in 2+1 dimensions is orthogonal to surfaces conformal to circles.}}
	\label{fig:twoplusone}
\end{figure}

 In order to define the infall momentum, we introduce a vector field $C^a$ on $\Sigma$ which selects  the appropriate  `radial' momentum component, namely we define the $C$-infall momentum
to be 
\be\label{infallc}
P_C [\Sigma] = \int_{{\Sigma}} {\cal P}_C 
\;,
\ee
where 
$
{\cal P}_C =  {\cal P}_a \,C^a 
$.
Multiplying (\ref{mc}) by $C^b$ and integrating by parts we obtain the following expression for the $C$-infall momentum: 
\be\label{mcin}
\int_\Sigma {\cal P}_C = -{1\over 8\pi G} \int_{\pt \Sigma } \dd S^a \,K_{ab}\, C^b  +{1\over 8\pi G} \int_{\Sigma} K^{ab} \,\nabla_a C_b\;.
\ee
Demanding that the boundary integral computes $\dd {
	\cal C}/\dd t$, according to \eqref{bint} and \eqref{bu},  fixes the\textit{ ansatz }for the infall current: 
\be\label{currentC}
J_a = -{b \over d-1} \,K_{ab}\,C^b\;,
\ee
with boundary condition 
\be\label{bcon}
-{b \over d-1} \,K_{ab} \,C^b \Big |_{\pt \Sigma} = (\pt_t)_a \Big |_{\pt \Sigma} \;. 
\ee
Incidentally,  the occurrence of the `box size'  in \eqref{bcon} shows that $b$ can be interpreted as setting the dimensions of the infall vector field. 
Finally, defining the  `remainder term' by the expression
\be \label{creminder}
R_C \,[\Sigma] = -{1\over 8\pi G} \int_\Sigma K^{ab} \,\nabla_a C_b\;,
\ee
we have established a PVC correspondence
\be\label{PVC}
{\dd  {\cal C}\over \dd t} = P_C \,[\Sigma] + R_{C} \,[\Sigma]\;.
\ee
This shows that part of the complexity rate at time $t$ can always be attributed to momentum flow through $\Sigma_t$. This PVC correspondence becomes  `exact' when $C$ can be chosen so as to make the remainder term vanish: 
\be\label{exactpvc}
{\dd {\cal C} \over \dd t} = P_C \,[\Sigma_t] = -\int_{\Sigma_t} N^\mu \,T_{\mu\nu} \,C^\nu\;, 
\ee
with $C^\mu \equiv e^\mu_a \,C^a$. As shown in \cite{prd}, a sufficient condition for this to happen is for   $\Sigma_t$ to be topologically and conformally trivial. This includes  single-boundary states with either spherical symmetry or in  $(2+1)$ dimensions (cf. Figure \ref{fig:twoplusone}), for which one can pick $C$ as a radial conformal Killing vector  satisfying 
\be\label{ckv}
\nabla_{(a} C_{b)} - {1\over d} \,h_{ab}\, \nabla_c C^a =0\;.
\ee

However, the generic state should  have a non-vanishing remainder, such as any pure gravitational wave configuration with ${\cal P}_a=0$.\footnote{It was pointed out in \cite{prd} that
non-trivial topology of $\Sigma_t$ may also force a non-zero remainder, even for states of high symmetry.} The outstanding question we want to address is whether there exist a generalization of the notion of `infall momentum' which works in these more general situations.        

\subsection{Generalized PVC from the Codazzi Equation}

The key idea in obtaining a generalized PVC relation is to replace the momentum constraint by a more general  starting point. The natural candidate is the Codazzi equation (cf. \cite{Poisson}) 
\be\label{codazzi}
\nabla_{c}\,K_{ab}\, - \,\nabla_{b}\,K_{ac}\, = \, N^\mu\,R_{\mu\nu\sigma\rho}\,e^{\nu}_{a}\,e^{\sigma}_{b}\,e^{\rho}_{c}\;,
\ee
since the momentum constraint \eqref{mc} is contained in its trace. Notice however that \eqref{codazzi}  involves a projection of the full Riemann tensor instead of simply the Ricci tensor. Therefore,   the difference between \eqref{codazzi} and \eqref{mc} is proportional to the Weyl tensor $W_{\mu\nu\rho\sigma}$. Since gravitational waves are precisely characterized by a non-vanishing Weyl tensor, the Codazzi equation  has the right ingredients for the kind of  generalization that we are seeking.

Following the same steps of the previous section, we  want to integrate \eqref{codazzi} over the extremal surface $\Sigma$. In doing so, we need to contract the three free indices with an `infall'  rank-3 tensor field, $M^{abc}$, enjoying  the same symmetry properties as the Codazzi equation, namely antisymmetry in the last two indices, $M^{abc} = - M^{acb}$, and the cyclic  identity $M^{abc}+M^{bca}+M^{cab} = 0$. 
Using  these symmetry properties, the Ricci decomposition of the Riemann tensor and Einstein's equation, we can rewrite \eqref{codazzi} as 
\be\label{codazzicontracted}
-M^{abc}\,\nabla_{c}\,K_{ab}\, = \,\dfrac{8\pi G}{d-1}\,\mathcal{P}_C\,+\,\dfrac{1}{2}\,{\cal W}_M \;,
\ee
where ${\cal P}_C = {\cal P}_a C^a$, and the $C$-field 
\be\label{cind}
C^b = h_{ac} \,M^{abc}
\ee 
is an infall vector field induced by the infall  tensor field. The  density 
\be\label{defw}
{\cal W}_M  = -N^\mu W_{\mu \nu\rho\sigma} \,e^\nu_a \,e^\rho_b\,e^\sigma_c \; M^{abc} 
\ee
is the contraction of the  infall tensor field with  the pulled-back, projected Weyl tensor. 
Integrating now by parts we find
\be\label{intcond}
-\int_{\partial \Sigma}\,\text{d}S_a\,K_{bc}\,M^{bca}\, = \, \dfrac{8\pi G}{d-1}\,\int_{\Sigma}\,\mathcal{P}_C\,+\,\dfrac{1}{2}\int_{\Sigma}\,\mathcal{W}_M\, - \int_\Sigma \,K_{ab} \,\nabla_c \,M^{abc} \;.
\ee

An interpretation of the  left-hand side of \eqref{intcond} as computing $\dd{\mathcal{C}}/\dd t$ requires that we set 
\be\label{currentM}
J^a\, = \, - b\, K_{bc}\,M^{bca}\;
\ee
in the notation of \eqref{bu}, with  the boundary condition 
\be\label{bcondm}
-b \,K_{ab} \,M^{abc} \Big |_{\pt \Sigma} = (\pt_t)^c \Big |_{\pt \Sigma}
\;.
\ee
Finally, defining the remainder term 
\be\label{rem}
R_M [\Sigma] = -{d-1 \over 8\pi G} \int_\Sigma K_{ab} \,\nabla_c \, M^{abc} \;,
\ee
and the integrated  `Weyl momentum'
\be\label{weylm}
W_M [\Sigma] = {d-1 \over 16\pi G} \int_\Sigma {\cal W}_M \;, 
\ee
we readily obtain a tensor generalization of \eqref{PVC} 
\be\label{MpvcR}
{\dd {\cal C} \over \dd t} = P_C [\Sigma_t] + W_M [\Sigma_t] + R_M [\Sigma_t]
\;.
\ee

This equation is completely general, under the assumption that the boundary condition \eqref{bcondm} is satisfied. It generalizes \eqref{PVC} in two ways. First, it contains information about bulk dynamics that goes beyond the mere initial-value constraints of GR, since it stems from the Codazzi equation. Second, it requires a generalization of the notion of `infall' vector field into a tensor infall field with many more components. An immediate consequence of this increase in degrees of freedom materializes when we consider sufficient conditions for the remainder to vanish. 

If the infall tensor $M^{abc}$ can be further restricted so that the remainder \eqref{rem} vanishes, that is to say, if $M^{abc}$ can be chosen satisfying  the conditions:
\be\label{strc}
 -b \,K_{ab} \,M^{abc} \Big |_{\pt \Sigma} = (\pt_t)^c \Big |_{\pt \Sigma}\;, \qquad K_{ab}\,\nabla_c M^{abc} =0\;,
\ee
then  we obtain the remainder-free, generalized PVC correspondence
\be\label{genPVC}
{\dd {\cal C} \over \dd t} = P_C [\Sigma] + W_M [\Sigma]\;.
\ee
Its main novelty compared to the restricted PVC \eqref{exactpvc} is the presence of a purely  gravitational contribution to the infall momentum, $W_M [\Sigma]$,  formally depending on    the Weyl curvature. In particular, states with no `clumping' matter,  having $ P_C =0$, still pick the Weyl contribution to the complexity rate. This we will see explicitly in the next section, in a particular example. 

We now collect a few  observations regarding our proposed generalization of the PVC correspondence.

\begin{itemize}

	\item
	The generalized relations \eqref{MpvcR} and \eqref{genPVC} reduce to the `restricted' ones \eqref{PVC} and \eqref{exactpvc}  when the infall tensor field admits the factorized {\it ansatz} 
\be\label{factor}
M^{abc}\, = \, \dfrac{1}{d-1}\,\left(h^{ac}\,C^b\,-\,h^{ab}C^c\right)\;, 
\ee
in terms of some `infall' vector field $C^a$. For sufficiently localized bulk states, this {\it ansatz} can be used to solve the box boundary condition \eqref{bcondm} when we place the box at infinity in an asymptotically flat or AdS spacetime.  As explained in  \cite{prd} for the AdS case, the solution involves a $C$-field with asymptotic behavior 
\be\label{cfielda}
C \rightarrow -{1\over b} \,r(y) \, \pt_y 
\ee
as $r\rightarrow \infty$. In this expression, the radial coordinate $y$ and the function $r(y)$ are such that the asymptotic approximation to the induced metric is given by 
\be\label{asympin}
\dd s^2_{\Sigma} \rightarrow \dd y^2 + r^2 (y) \,\dd \Omega_{d-1}^2\;.
\ee
In  Appendix \ref{sec:appendixB} we extend this result to the asymptotically flat case. 
Conceptually, the combination of \eqref{factor} and \eqref{cfielda} shows that, asymptotically,  the `infall' interpretations of $M^{abc}$ and $C^a$ reduce to one another.

\item 	

We can look for sufficient conditions  for the vanishing of the remainder \eqref{rem},  which would generalize the conformal Killing condition  \eqref{ckv}.  Given that $\Sigma_t$ is extremal, with $K=0$, the remainder \eqref{rem} vanishes if  the  symmetrized divergence of $M^{abc}$   is a conformal rescaling, $\nabla_c \,M^{(ab)c} = \Phi \,h^{ab}$ for some scalar function $\Phi$ on $\Sigma$. Taking traces, we can compute $\Phi$ and obtain the equivalent trace-free transversality condition 
\be\label{transv}
\nabla_c \left( M^{(ab)c} - {1\over d}\, h^{ab} \,h_{ef} \,M^{efc} \right) =0\;.
\ee
We can expect  that 
finding transverse tensors satisfying \eqref{transv} on $\Sigma$ should be easier than finding conformal Killing vectors on $\Sigma$, simply as a consequence of the existence of many more degrees of freedom in $M^{(ab)c}$. Within the factorized {\it ansatz} \eqref{factor}, the $M$-transversality condition \eqref{transv} reduces to \eqref{ckv}, giving  the most general solution of \eqref{transv}  in spherically symmetric spacetimes or any $(2+1)$-dimensional solution, precisely those cases in which the restricted PVC is exact (cf. Appendix \ref{sec:appendixA}). 

\item

The generalized PVC relation presented here is bound to suffer from  similar topological obstructions as the restricted PVC. For an eternal black hole state with spherical symmetry, extremal hypersurfaces $\Sigma_t$ have two disconnected boundaries, the infall tensor field satisfies the factorized {\it ansatz} \eqref{factor} and
the generalized `Weyl momentum' still vanishes when evaluated on the extremal surfaces. This means that the result of \cite{prd} still applies, i.e.   the rate of complexity growth with forward time variables on both sides satisfies \eqref{MpvcR} with a non-zero remainder term $R_M [\Sigma_t]$, which in this example happens to coincide with  \eqref{creminder}. Therefore, the example of the eternal black hole implies that an exact PVC relation with vanishing remainder will always require some topological assumptions about the extremal surfaces.  

\end{itemize}

\section{An Explicit Check of the Generalized PVC}
\label{sec:PPWaves}

In this section we present a detailed verification of  \eqref{strc} and   \eqref{genPVC} on a non-trivial exact solution of Einstein's equations: a gravitational pp-wave. 
Gravitational waves are usually discussed in the context of a perturbative expansion around a flat background. In such cases, one usually makes the \textit{ansatz}
\begin{equation}
	g_{\mu \nu} = \eta_{\mu \nu} + h_{\mu \nu}\, ,
\end{equation}
where $ h_{\mu \nu}$ is treated as a small perturbation and its dynamics is determined by the linearized GR theory. The standard analysis shows that gravitational waves are transversely polarized, that is to say, in a suitable coordinate frame,   a wave travelling in the $z$ direction will only distort the metric in the transverse directions $y^i=\lbrace x,y \rbrace$, yielding a metric of the form
\begin{equation}
	\label{perturbativeGW}
	\dd s^2 = -\dd u\dd v +(\delta_{ij} + h_{ij}(u))\dd y^i \dd y^j,
\end{equation}
where we have defined the light-like coordinates $u=t-z$ and $v=t+z$. 

Exact non-perturbative gravitational wave solutions to the Einstein equations are highly idealized objects, but nevertheless exist and might be useful for pedagogical purposes. One way of defining them is through a direct generalization of \eqref{perturbativeGW}, dropping the requirement that $h_{\mu \nu}$ is small and considering solutions of the form
\begin{equation}
	\label{Rosen}
	\dd s^2 = -\dd u\dd v +g_{ij}(u)\dd y^i \dd y^j\, ,
\end{equation}
known as the Rosen form of a gravitational plane wave with parallel propagation (PP) solution. In order to be able to perform some calculations, we will sacrifice some generality by sticking to the following particular\textit{ ansatz }(cf. \cite{MTW, Bondi, EhlersKundt})
\begin{equation}
	\label{RosenAnsatz}
	\dd s^2 = -\dd u\dd v +L^2(u)\left( e^{2 \beta(u)} \dd x^2 + e^{-2 \beta(u)} \dd y^2 \right)\, ,
\end{equation}
where the functions $L(u)$ and $\beta(u)$  are to be determined by the Einstein equations. As the solution represents a null wave by construction, the only component of the Ricci tensor that is excited is
\begin{equation}
	R_{uu} = -2 L^{-1} \left( L'' + (\beta')^2 L \right)\, ,
\end{equation}
where the primes stand for $\dd / \dd u$. Demanding a purely gravitational solution therefore requires
\begin{equation}
	\label{Ricciflatcondition}
	L'' + (\beta')^2 L =0\, 
\end{equation}
to hold. Once this condition is satisfied, the manifold still possess a non-trivial Riemann curvature, with the non-zero elements given by
\begin{eqnarray}
	\label{Riemann}
	R_{uxux}&=&-e^{2 \beta}L\,(2L'\beta'+L\beta'')\,, \\
	R_{uyuy}&=&-e^{-2 \beta}L\,(2L'\beta'+L\beta'') \, \nonumber .
\end{eqnarray}

\begin{figure}[h]
	\begin{center}
		\includegraphics[width=8cm]{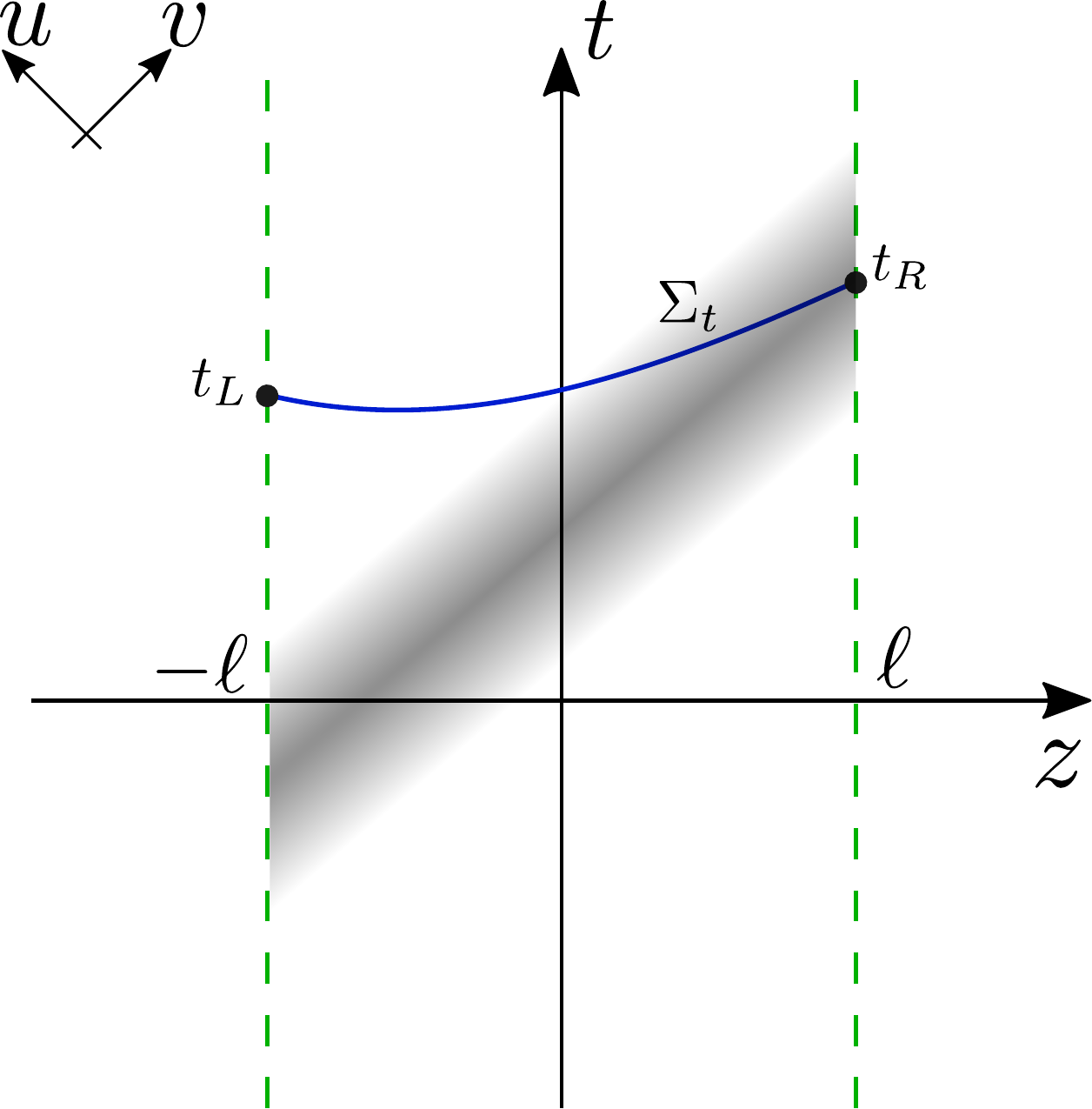}
		\caption{\textit{Schematic representation of the extremal surface for a pp-wave pulse traversing the box of size $2\ell$ from left to right. The transverse ${\bf R}^2$ plane is not shown.}}
	\end{center}
	\label{fig:PP}
\end{figure}

We shall check the generalized PVC correspondence for a region $X$ defined by the slice delimited by $-\ell \leq z \leq \ell$. The boundary $\pt X$ has left and right disconnected components $z = \pm \ell$. The extremal surface has the form $\Sigma = \gamma \times {\bf R}^2$, where the curve $\gamma$ is anchored at $\pt X$ on times $t_L$ and $t_R$ (see Figure \ref{fig:PP}). We will fix the conventional normalization of the complexity by choosing the  `box size' $b$ to equal the coordinate edge of $X$   in the $z$ direction, namely  we set $b= \ell$.

In checking \eqref{genPVC}, we shall factor the formally infinite transverse volume of the ${\bf R}^2 $ component, namely we aim to compute both sides of 
\be\label{ddd}
{\dd V_\gamma \over \dd t} = {\ell \over 2V[{\bf R}^2]} \int_\Sigma {\cal W}_M \;,
\ee
where $V_\gamma = V[\Sigma] / V[{\bf R}^2]$ is the longitudinal volume of a symmetrically anchored curve, with $t_L = t_R = t$. 

\subsubsection*{Computation of the Extremal Volume} 
\label{sec:onshell}

Picking an arbitrary parametrization of $\gamma$, the longitudinal volume is given by 
\begin{equation}
	V_\gamma = \int\limits_{\lambda_L}^{\lambda_R} \dd \lambda \sqrt{-L(u)^4\dot{v}\dot{u}} \, ,
\end{equation}
where the dot stands for $\dd / \dd \lambda$ and, as we see, the dependence on the function $\beta(u)$ drops off the determinant of the induced metric on the slice, significantly simplifying the extremalization problem. Furthermore, we may choose the coordinate $u$ itself as our $\lambda$ parameter, yielding the effective lagrangian
\begin{equation}
	\mathcal{L} = \sqrt{-L(u)^4 \dot{v}}\, .
\end{equation}
As the action does not depend explicitly on $v$ we can obtain a first integral for the Euler-Lagrange equations from the canonical momentum associated to $v$
\begin{equation}
	\label{pdef}
	p = \dfrac{\partial \mathcal{L}}{\partial \dot{v}} = \dfrac{L(u)^2}{2 \sqrt{-\dot{v}}}\, .
\end{equation}
Feeding this back into the action, we get that the on-shell volume will be
\begin{equation}
	V_\gamma = \dfrac{1}{2p}\int\limits_{u_R}^{u_L} \dd u \; L(u)^4 = 2p \left[ v(u)\right]^{u_L}_{u_R}\, ,
\end{equation}
meaning that with this choice of parameters $v(u)$ itself measures (up to a multiplicative constant) the volume along the slice. Integrating \eqref{pdef} we get 
\begin{equation}
	v(u) = -\dfrac{1}{4p^2}\int \dd u \; L(u)^4 +c\, ,
\end{equation}
and we can fix the value of the constants $p$ and $c$ by imposing the boundary conditions $v(u_{L,R})=v_{L,R}$. Solving for $p$ we can write the volume purely in terms of the unknown function $L(u)$ and the boundary values
\begin{equation}
	V_\gamma =\sqrt{ (v_R-v_L) \int\limits_{u_R}^{u_L} \dd u \; L(u)^4}\, ,
\end{equation}
which is a completely general expression for arbitrary boundaries. For the simpler setup $z_R=-z_L=\ell $ and considering also a symmetric evolution $t_R=t_L=t$,  we may calculate the rate of growth, which gives
\begin{eqnarray}
	\label{Vrate}
	\dfrac{\dd V_\gamma } {\dd t} &=& \dfrac{1}{4p}\left[L(u_L)^4-L(u_R)^4 \right] \, .
\end{eqnarray}

As a check, we see that the trivial flat solution ($L(u)=1$ and $\beta(u)=0$) gives the expected behaviour, i.e. an extremal surface given by the straight line
$
	v(u) = -u+2t$, 
corresponding to the fixed-$t$ surface. The volume of this surface is simply
	$V= 2 \ell $,
and of course its growth rate vanishes. 

Notice that any other non-trivial profile for $L(u)$ enjoying the symmetry $L(u_L)^4=L(u_R)^4$ will have vanishing rate as well. This is for example the case of pp-waves that have a finite extension in the $u$ direction (e.g. a compactly supported pulse). The volume may change as the wave enters or leaves the region but will stay constant if the metric at the boundaries remains flat.

\subsubsection*{Computation of the Gravitational Infall Momentum}

After solving explicitly the variational problem for the pp-wave spacetime \eqref{RosenAnsatz}, we will now  evaluate the gravitational infall momentum  induced by the Weyl tensor. 
To begin with, recall that the embedding functions $X^\mu = \lbrace u, v(u), x,y  \rbrace$ defining the surface are given by
\begin{equation}
	\label{maxsurface}
	v(u) = -\dfrac{1}{4p^2}\int \dd u \; L(u)^4 +c \, ,
\end{equation}
so, choosing the coordinates on $\Sigma$ to be $y^a = \lbrace u,x,y \rbrace$ we can readily calculate the tangent vectors
\begin{equation}
	e_a^\mu = \dfrac{\p X^{\mu}}{\p y^a} =
	\left(
	\begin{array}{cccc}
		1 & -\frac{L(u) ^4}{4 p^2} & 0 & 0 \\
		0 & 0 & 1 & 0 \\
		0 & 0 & 0 & 1 \\
	\end{array}
	\right)\, ,
\end{equation}
and the induced metric on the slice
\begin{equation}
	h_{ab} = g_{\mu \nu} e^\mu_a e^\nu_b =
	\left(
	\begin{array}{ccc}
		\frac{L (u)^4}{4 p^2} & 0 & 0 \\
		0 & L (u)^2 e^{2 \beta  (u)} & 0 \\
		0 & 0 & L (u)^2 e^{-2 \beta  (u)} \\
	\end{array}
	\right)\, .
\end{equation}
The timelike normal vector is
\begin{equation}
	N^\mu = \left(\frac{2 p}{L(u)^2},\frac{L(u)^2}{2 p},0,0\right)\, ,
\end{equation}
which allows us to compute the extrinsic curvature of the surface
\begin{equation}
	K_{ab}= \nabla_\mu N_\nu e^\mu_a e^\nu_b = \left(
	\begin{array}{ccc}
		-\frac{L L'}{p} & 0 & 0 \\
		0 & \frac{2 p\, e^{2 \beta } \left(L'+L \beta '\right)}{L} & 0 \\
		0 & 0 & \frac{2 p \,e^{-2 \beta } \left(L'-L \beta '\right)}{L} \\
	\end{array}
	\right) \; ,
\end{equation}
which is of course traceless ($K=0$) due to the extremal character of $\Sigma$.

We now have all the ingredients at hand to determine the infall tensor field  satisfying the differential equation and boundary conditions
\begin{eqnarray}
	K_{ab} \nabla_c M^{abc}&=&0\, , \label{differentialcondition}\\
	-\ell\,K_{ab}M^{abc}\Big|_{\p \Sigma}&=& (\pt_t)^\mu \, e^c_\mu\, \Big|_{\p \Sigma} \,. \label{boundarycondition}
\end{eqnarray}
The dynamical equation gives
\begin{eqnarray}
	& &e^{2 \beta } \left(L'+L \beta' \right) \left( M^{xxu} \left(3 L'+L \beta' \right)+L M'^{xxu} \right) \\&+& e^{-2 \beta } \left(L'-L \beta' \right) \left(M^{yyu}  \left(3 L'-L \beta' \right)+L M'^{yyu} \right)=0\, , \nonumber
\end{eqnarray}
from which it is easy to find a particular solution given by
\begin{eqnarray}
	M^{yyu} &=& A \, e^{\beta}L^{-3}\, , \label{solutionMuyy} \\
	M^{xxu} &=& B \, e^{-\beta}L^{-3}\, .  \label{solutionMuxx}
\end{eqnarray}
$A$ and $B$ are integration constants to be fixed by the boundary conditions \eqref{boundarycondition} which become
\begin{equation}
	\frac{e^{-\beta} \left(2 p L \beta ' \left(A-B e^{2 \beta }\right)-2 p L' \left(A+B e^{2 \beta }\right)\right)}{L^4}\Bigg|_{u=u_L, u_R}=\dfrac{1}{2\ell} -\dfrac{2p^2}{\ell L^4}\Bigg|_{u=u_L, u_R} \, .
\end{equation}
Actually, since \eqref{boundarycondition} is a vectorial equation, two more scalar equations impose certain algebraic conditions on other tensor components of $M^{abc}$. These components  will nevertheless  be annihilated upon contraction with the Weyl tensor \eqref{Riemann} in the generalized PVC formula, so we will  not calculate them here.
Solving for the constants $A$ and $B$  we get
\begin{eqnarray}
	\ell\,A&=&\frac{e^{\beta_L+\beta_R} \left(e^{\beta_R} \left(4p^2-L_L^4\right) \left(L'_R+L_R \beta'_R\right)-e^{\beta_L} \left(4 p^2-L_R^4\right) \left(L'_L+L_L \beta'_L\right)\right)}{4 p e^{2 \beta_R} \left(L'_L-L_L \beta'_L\right) \left(L'_R+L_R \beta'_R\right)-4 p e^{2 \beta_L} \left(L'_L+L_L \beta'_L\right) \left(L'_R-L_R \beta'_R\right)}\, , \\
	\ell\,B&=&\frac{e^{\beta_R} \left(4 p^2-L_R^4\right) \left(L'_L-L_L \beta'_L\right)-e^{\beta_L} \left(4 p^2-L_L^4\right) \left(L'_R-L_R \beta'_R\right)}{4 p e^{2 \beta_R} \left(L'_L-L_L \beta'_L\right) \left(L'_R+L_R \beta'_R\right)-4 p e^{2 \beta_L} \left(L'_L+L_L \beta'_L\right) \left(L'_R-L_R \beta'_R\right)}\, , \label{Bconstant}
\end{eqnarray}
where we used the notation $L_{L,R} = L(u_{L,R})$. Embedding the tensor into the four-dimensional spacetime $M^{\nu \rho \sigma} = M^{abc}e_{a}^\nu e_{b}^\rho e_{a}^\sigma,$ and contracting it with the Weyl and normal vectors we can finally obtain the longitudinal portion of the gravitational infall momentum 

\begin{eqnarray}
	\label{SOMFrate}
	{\ell \over 2V[{\bf R}^2]}  \int_\Sigma {\cal W}_M &=&  -{\ell \over 2V[{\bf R}^2]}  \int_\Sigma N^\mu \,W_{\mu\nu\rho\sigma} \,M^{\nu\rho\sigma}  \nonumber \\ 
	&=&  \dfrac{\ell}{2}\int\limits^{u_L}_{u_R} \dd u \left(Ae^{-\beta (u)}  -B e^{ \beta (u)}\right) \left(2 L'(u) \beta '(u)+L(u) \beta ''(u)\right) \label{SOMFrateintegral}\\ &=& \dfrac{1}{4p} \left[ L(u_L)^4 - L(u_R)^4 \right] \, ,  \nonumber
\end{eqnarray}
which of course is exactly \eqref{Vrate}, the same result that we obtained with the direct extremalization procedure. The vanishing of the total gravitational infall momentum for a perfectly contained pulse, having $L(u_L) = L(u_R)$,  is consistent with the idea that the pp-wave is `infalling' from the point of view of the left boundary, but is equally `outfalling' from the point of view of the right boundary.\\

\subsubsection*{Observations}

We conclude this section with a few observations regarding these explicit computations.

\begin{itemize}

	\item
	The pp-wave example illustrates that a form of the PVC correspondence holds for pure gravitational waves in asymptotically flat spacetime but, more emphatically,  the `box'  can have a finite size, defined by some conventional coordinate condition. If the $M$-tensor can be defined as satisfying \eqref{strc}, the generalized PVC holds with no need to make special physical arrangements to define the walls of the box. In this sense we can say that the generalized PVC correspondence is a quasilocal property of the bulk dynamics. 
	\item 
	Even though we chose a purely gravitational solution in order to maximize the differences with our previous analysis  for collapsing matter solutions (cf. \cite{ushell}), notice that most of the details go through even after dropping the Ricci-flatness condition \eqref{Ricciflatcondition}. In such case, the\textit{ ansatz }\eqref{RosenAnsatz} describes in general a mixture of a gravitational and a (null) matter pp-wave with an energy-momentum tensor given by 
	\begin{eqnarray}
		T_{\mu\nu} = -\dfrac{2}{8 \pi G}  L^{-1} \left( L'' + (\beta')^2 L \right)\delta^u_\mu\delta^u_\nu \, .
	\end{eqnarray}
	Of course, for this model to describe real matter one should ask $T_{\mu\nu}$ to satisfy certain null energy conditions, which in turn will impose some restrictions on the functions $L(u)$ and $ \beta(u)$. At any rate, our observation here is that 
	this non-trivial Ricci curvature does not change any of our analysis from \eqref{maxsurface} to \eqref{Bconstant} since all quantities on the slice depend only on first order derivatives of the metric, yielding formally identical results for the $M^{\mu\nu\sigma}$ tensor field. The total volume variation does however pick an additional term from the matter momentum
	\begin{eqnarray}
		\label{VWplusP}
		\dfrac{\dd V_\gamma}{\dd t} = {\ell \over V[{\bf R}^2]} \left[ \dfrac{1}{2}\int_\Sigma {\cal W}_M  - \int_\Sigma N^\mu T_{\mu \nu} C^\nu \right]\;,
	\end{eqnarray}
	where the `infall vector field' $C^\mu$ defined in \eqref{cind} can be easily obtained from our solution for $M^{abc}$ yielding
	\begin{equation}
		C^a = \ell^{-1}L^{-1}(u)\left( A e^{-\beta (u)} + B e^{\beta (u)}\right)  \delta_u^a\, .\\
	\end{equation}
	As it can be readily checked, the sum of the two contributions in \eqref{VWplusP} nicely recovers the correct result \eqref{Vrate} again without any additional contribution. The technical reason behind this relies on the existence of the total derivative
	\begin{eqnarray}
		\dfrac{\dd F(u)}{\dd u} &=&  \sqrt{h} \left[\dfrac{8\pi G}{d-1}\mathcal{P}_C + \dfrac{1}{2}\mathcal{W}_M \right] \, ,\\
		F(u) &=& -\left(Ae^{-\beta} + B e^\beta\right)L'+\left(Ae^{-\beta} - B e^\beta\right)L\beta' \, ,
	\end{eqnarray}
	which allows us to perform the integral over $\Sigma$ for arbitrary functions $L(u)$ and $\beta(u)$, including the Ricci and conformally flat solutions as well as any generic mixture of gravitational and matter waves.

	\item  
	
	As it is well known, the very concept of local energy and momentum becomes ill-defined when we try to adapt it to gravity itself, where only perturbative notions of an approximate energy-momentum pseudotensor $t_{\mu\nu}$ have been proposed (cf. \cite{Einsteintensor,Landau, AbbottDeser}). For that reason we do not expect to find a clear general relation between complexity growth and energy inflow for pure gravity solutions. We find however amusing that our example admits an interpretation along these lines. In particular, it is possible to find a `gravitational inflow vector' $\tilde{C}^\mu$ which allows us to re-write the integrand in \eqref{SOMFrateintegral} in a similar fashion as the matter piece, i.e.
	\begin{equation}
		W_{\mu \nu \rho \sigma} M^{\nu \rho \sigma} = {\tilde t}_{\mu\nu}\tilde{C}^\nu
	\end{equation}
	with 
	\begin{eqnarray}
		\tilde{C}^a = \ell^{-1}L^{-1}(u)\left( A e^{-\beta (u)} - B e^{\beta (u)}\right)  \delta_u^a\, ,\\
		{\tilde  t}^{}_{\mu\nu}= 2\left(\dfrac{2L'\beta'+ L\beta''}{L} \right)\delta_\mu^u \delta_\nu^u \, ,
	\end{eqnarray}
	where we can identify ${\tilde t}_{\mu \nu}$ as the `square root' \footnote{See \cite{BonillaSenovilla} for a proper definition of this object} of the Bel-Robinson tensor (cf. \cite{SenovillaSuperEnergy}), an object that is constructed purely from the Weyl tensor and satisfies the dominant property ${\tilde t}_{\mu \nu} u^\mu u^\nu \geq 0$ for any future-pointing vector $u^\mu$ (cf. \cite{SenovillaCausal}). It would be interesting to investigate whether this formal analogy still holds beyond the particular example at hand.

\end{itemize}

\section{Discussion}
\label{sec:conclusions}

In this paper we have presented a further generalization of the idea that certain holographic complexity/momentum correspondences are largely implicit in the dynamics of Einstein gravity. The fundamental working assumption is that   complexity is computed by   extremal codimension-one  volumes in the bulk, i.e. we adopt the Complexity=Volume prescription \cite{SusskindEntnotEnough}.

The main realization is that the  `matter PVC', presented in \cite{prd} as a consequence of the momentum constraint of GR, admits a nontrivial generalization into a fully gravitational PVC stemming from the Codazzi equation: 
\be\label{otravezgen}
{\dd {\cal C} \over \dd t} = P_C [\Sigma_t] + W_M [\Sigma_t]\;.
\ee
The main novelty of this PVC relation is the occurrence of a new contribution to the `rate of gravitational clumping', measured by an appropriate flux of the Weyl tensor.  A crucial technical  ingredient  is the generalization of the notion of `infall vector field' $C^a$ into a rank-3  `infall tensor field'  $M^{abc}$ with the same symmetry structure as the Codazzi equation itself.  In order for \eqref{otravezgen} to be true, $M^{abc}$ must be chosen to satisfy the equation
\be\label{remzero}
K_{ab} \, \nabla_c M^{abc} =0
\ee
throughout $\Sigma_t$, with boundary conditions \eqref{bcondm}. We have explicitly checked that these requirements can be met  in an exact pure-gravity pp-wave solution of Einstein's equations. 

This result poses a number of interesting questions. While \eqref{otravezgen} is certainly more general than \eqref{exactpvc}, we are still lacking a more precise physical interpretation of the Weyl-momentum $W_M [\Sigma_t]$. It would be interesting to explore possible connections to pseudo-local energy notions based on the Bel--Robinson tensor, as suggested at the end of the previous section. Further elucidation along these lines will follow form a careful analysis of weak-field expansions around the asymptotic factorized {\it ansatz} \eqref{factor}. 

The generality of the PVC relations presented in this paper should allow us to tackle the case of non-trivial boundary dynamics in AdS/CFT examples. This includes the VC-complexity of  `cosmological' constructions driven by time-dependent states in the CFT, as in \cite{barbonrabino}, and `boundary gravitons' in $2+1$ dimensions \cite{brownhenneaux, CaputaMagan, NY, Flory}. It would be interesting to study the detailed solutions of \eqref{strc} that arise in these situations, where the PVC relation is expected to contain  additional  `boundary' contributions beyond the bulk infall momenta.

At a purely mathematical level, it would be interesting to delimitate the reach of sufficient conditions such as the symmetrized transversality condition on the infall tensor \eqref{transv}. The answer is guaranteed to be nontrivial, for at least two reasons. First, the explicit solution we have found for $M^{abc}$  in the pp-wave example does {\it not } satisfy \eqref{transv}. Therefore, we know that in cases that are sufficiently far from the factorized {\it ansatz} \eqref{factor}, the transversality condition is too strong. Second, even when the factorized {\it ansatz} works and \eqref{transv} reduces to the conformal Killing condition, topological obstructions can prevent the remainder from vanishing.

We end with a digression on the more general significance of PVC relations like \eqref{otravezgen}.  First of all, our proposed PVCs are tailor-made for the VC prescription. By now, a plethora of different complexity proposals exist \cite{AC1,AC2, NullBoundariesAC, Fishler, Takayanagi2, Flory, CaputaMagan, NY} and it would be interesting to see  if analogous momentum/complexity correspondences can be formulated. When addressing this question, one should keep in mind that subtly different notions of complexity may exist in the boundary description. As a simple  example of this fact, we can consider operator K-complexity \cite{Altman, Barbon, Rabinovici, Swingle}, which is conceptually different from circuit complexity, yet it shows analogous `phenomenology' in certain situations.

At any rate,  we know that extremal spatial volumes parametrized by codimension-one boundary data are interesting quantities in any putative holographic description. Whether they are literally related to some sort of computational complexity is an open question, but it is certain that there exists a  notion of `volume complexity' induced from the bulk description. In this context, one can imagine using the PVC formula as a basis for its elucidation. Since the right hand side of \eqref{otravezgen} is a local bulk integral, we can expect that a sufficiently powerful prescription of bulk operator reconstruction can be used to give an operational definition of $\dd {\cal C}/ \dd t$ in the dual holographic picture (CFT or otherwise). A further integration determines the `volume complexity' up to a constant, mimicking the strategy followed before to determine the Newtonian limit of the complexity in equations \eqref{clump} and \eqref{pcc}. In this context, it becomes interesting to investigate the relation between the PVC correspondence and other structural properties of holographic complexity, such as \cite{FirstLaw,FirstLaw2,Secondlaw}. We hope to come back to these questions in  future investigations.

\subsection*{Acknowledgements}

We would like to thank C. Gomez, J. F. Pedraza, A. Russo, A. Svesko and Z. Weller-Davies  for discussions. This work is partially supported by the Spanish Research Agency (Agencia Estatal de Investigaci\'on) through the grants IFT Centro de Excelencia Severo Ochoa SEV-2016-0597,  FPA2015-65480-P and PGC2018-095976-B-C21. The work of J.M.G. is funded by grant FPA2017-84436-P from Ministerio de Economia y Competitividad, by Xunta de Galicia (Centro singular de investigaci\'on de Galicia acreditation 2019-2022) by ED431C 2017/07 by FEDER, by Eruopean Union ERDF, and by the "Mar\'ia de Maeztu" Units of Excellence program MDM-2016-0692 and the Spanish Research State Agency. The work of M.S. is funded by the FPU Grant FPU16/00639.

\appendix

\section{Recovering the Exact PVC for Special Cases}
\label{sec:appendixA}
\noindent

In this appendix, we show that the factorized \textit{ansatz} \eqref{factor} for the infal tensor $M^{abc}$ is the most general solution of the trace-free transversality condition \eqref{transv} for generic 2+1 dimensional spacetimes as well as for spherically symmetric solutions in higher dimensions. The boundary condition \eqref{bcondm} reduces to \eqref{bcon} for the $C$-field, which is now restricted to be a conformal Killing vector. With previous knowledge of the required asymptotic behavior for the $C$-field in AdS (cf. \cite{prd}), we also comment on how the generalized PVC reduces to the exact PVC for extremal volume slices anchored to the asymptotic boundary of AdS.

Let us first consider a generic spacetime in $2+1$ dimensions. The number of algebraically independent components of the infall tensor $M^{abc}$ for $d=2$ is 2, which precisely coincides with the number of trace-free transversality conditions \eqref{transv}. In order to solve them explicitly, we will choose coordinates locally on $\Sigma$ such that
\be\label{2dmetric}
\text{d}s^2_\Sigma \, = \, e^{2\omega(z,\bar{z})}\, \text{d}z\,\text{d}\bar{z}\;,
\ee
where $z=y+i\phi$, and $\omega(z,\bar{z})$ some real function. For the case of asymptotically AdS spacetimes, the metric \eqref{2dmetric} asymptotes the Poincaré disk metric $\omega\, \sim \,y$ as $y\rightarrow \infty$. We will suitably choose the two independent components of $M^{abc}$ to be the real and imaginary parts of $M^{zz\bar{z}}$ in these complex coordinates. The set of conditions \eqref{transv} becomes particularly simple in these coordinates
\bg
\partial_{\bar{z}}\,\left(e^{2\omega}\,M^{zz\bar{z}}\right)\, = \, 0 \;,\label{transv2d}\\
\partial_{z}\,\left(e^{2\omega}\,M^{\bar{z}\bar{z}z}\right)\, = \, 0\;.\label{transv2dcc}
\eg

It is straightforward to see that the most general solution of these equations is
\bg\label{m2+1}
M^{zz\bar{z}}(z,\bar{z})\, = \,2\, g(z)\,e^{-2\omega(z,\bar{z})}\;,
\eg
for some holomorphic function $g(z)$. The $C$-field obtained by taking the trace of this infall tensor is precisely $C^z \, = \, g(z)$. In two dimensions, every vector field of this form is locally a conformal Killing vector. The key observation is that this infall tensor field factorizes as $M^{abc}\, = \, h^{ac}\,C^b\,-\,h^{ab}\,C^c$. It then becomes clear the reason why the general solution of \eqref{transv} can be constructed from an infall $C$-field which is a conformal Killing vector. In AdS, the required asymptotic boundary condition is $C^y \sim -b^{-1}$ for the case of the unnormalized $y$ coordinate (cf. \cite{prd}) with $b = \ell_{\text{AdS}}$.  The unique holomorphic extension of this condition is to set $g(z) = -1$ throughout $\Sigma$. This way, we obtain the canonical $C$-field (cf. Figure \ref{fig:twoplusone}) which is orthogonal to the constant $y$ lines, inward pointing, and has a norm that depends on the point in question, $C^2\, = \, b^{-2}\,e^{2\omega}$. This infall field certainly coincides with the inward radial conformal Killing vector of the Poincaré disk. In fact, the Weyl-momentum vanishes in 2+1 dimensions as the Weyl tensor is exactly zero, which, together with the above definition of the $C$-field, shows how the generalized PVC reduces the exact PVC correspondence for any geometric state in 2+1 dimensions.

Let us now consider the case of a spherically symmetric spacetime in higher dimensions.  Assuming that $\Sigma$ inherits spherical symmetry, the induced metric can be written as
\be\label{sphericalmetric}
\text{d}s^2_\Sigma \, = \, \text{d}y^2 \, + r^2(y)\,\text{d}\Omega_{d-1}^2 \;,
\ee
where $y$ is an outward directed coordinate normal to the spheres. Moreover, it is natural to assume that the most generic infall tensor $M^{abc}$ is isotropic under $SO(d)$ (cf. \cite{Jeffreys}), up to possible terms that do not contribute to the boundary condition, and hence can be considered as pure gauge redundancies (tangent diffeomorphisms that die off asymptotically). The only irreducible isotropic rank-3 tensor is $\epsilon^{abc}$ for the case of $SO(3)$, but still this tensor does not lie in the same irreducible representation of $GL(d)$ as the infall tensor. Therefore, any isotropic $M^{abc}$ will necessarily be reducible into products of lower-rank tensors. The most general irreducible isotropic rank-2 tensor is of the form $f(y)\,h_{ab}$, where $h_{ab}$ is the spherically symmetric metric \eqref{sphericalmetric}. The most general isotropic vector is orthogonal to the spheres with an angle-independent norm, $C=  C(y)\,\partial_y$. With these building blocks in hand, there are two ways to construct an isotropic $M^{abc}$, i.e. from a rank-2 tensor and a vector $C^a\,h^{bc}$, or alternatively from three vectors $C_1^a\,C_2^b\,C_3^c$. The latter belongs to the totally symmetric representation of $GL(d)$, and hence it vanishes when projected into the representation of $M^{abc}$. Projecting the former provides then with the most general isotropic infall tensor
\be\label{isotropicgeneral}
M^{abc}\, = \, \dfrac{1}{d-1}\,\left(h^{ac}\,C^b\,-\,h^{ab}C^c\right)\;, 
\ee
which is again of the factorized form \eqref{factor}. In asymptotically AdS spacetimes, the required asymptotic boundary condition \eqref{boundarycondition} will be satisfied by the canonical inward radial $C$-field on the Poincaré ball $C = -b^{-1}\,r(y)\,\partial_y$ (cf. \cite{prd}). For any such factorized $M^{abc}$, the Weyl-momentum density will vanish due to tracelessness and antisymmetry of the Weyl tensor
\be
\mathcal{W}_M \, = \, -\dfrac{1}{d-1}\,N^\mu \, W_{\mu\nu\rho\sigma}\, \left(h^{\nu\sigma}\,C^\rho\,-\, h^{\nu\rho}\,C^\sigma\right)\, =\, \dfrac{2}{d-1}\,N^\mu \, W_{\mu\nu\rho\sigma}\,C^\sigma\, \left(g^{\nu\rho}\,+ N^\nu N^\rho\right)\, = \, 0\;,
\ee
which, together with the characterization of the $C$-field, leads to the exact PVC correspondence for any spherically symmetric normalizable state in $d+1$ dimensional asymptotically AdS spacetimes.

\section{Asymptotic Boundary Conditions}
\label{sec:appendixB}
\noindent

In this appendix, we extend the analysis of \cite{prd} of asymptotically AdS boundary conditions to include the asymptotically flat case. We elaborate on the asymptotic boundary conditions for the $C$-field and $M$-field that solve \eqref{bcon} and \eqref{bcondm} in both cases.

To start, we might adopt asymptotic coordinates in the vicinity of $\Sigma$ such that the metric reads
\be\label{ana} 
\text{d}s_X^2\, \rightarrow \, \dfrac{\text{d}r^2}{r^a}\,-\,r^a\,{\text{d}t^2}\,+\,r^2\,\gamma_{ij}(r,t,\theta)\,\text{d}\theta^{i}\text{d}\theta^j\hspace{5mm}\text{as}\hspace{3mm} r\,\rightarrow\, \infty \;.
\ee
where $a=2$ is the AdS$_{d+1}$ case and $a=0$ is the flat case.

Here, $r$ is an `asymptotically radial' coordinate which foliates $X$ by timelike codimension-one submanifolds $Y_r$. In the case of AdS, it corresponds to a Fefferman-Graham coordinate for a particular conformal frame. The angles $\theta^j$ parametrize the  $d$-dimensional intersection $S_r =  Y_r \cap \Sigma$, of spherical topology and induced metric proportional to $\gamma_{ij}$, which is itself asymptotic to a unit round sphere, up to normalizable corrections of order $1/r^d$. The time coordinate is chosen to be geodesic on $Y_r$ and orthogonal to $S_r$. \footnote{For $d=2$, we shall not consider here the possibility of large diffeomorphisms upon the {\it ansatz} \eqref{ana},  which would be required to discuss `boundary gravitons' in the sense of \cite{brownhenneaux}. }

The induced metric on $\Sigma$ can be written near the boundary as 
\be\label{induced}
ds^2_\Sigma \rightarrow dy^2 + r^2 (y) \,\gamma_{ij} (y, \theta) \,d\theta^i \,d\theta^j \;,
\ee
for some function $r(y)$ which asymptotically $r\sim a/2\,\sinh y\,+\,(1-a/2)\,y$ as $y\rightarrow \infty$.
This allows us to write the normal one-form  as 
$N_\Sigma = e^t_y \,dr - e^r_y \,dt$, and compute the extrinsic curvature $K_{ab} = e^\mu_a \,e^\nu_b \,\nabla_\mu \,N_\nu$. The relevant component turns out to be
$K_{yy}$ which, using the traceless character, $K=0$, may be evaluated as $K_{yy} = - r^{-2} \,\gamma^{ij} \,K_{ij}$. Explicitly 
\be\label{ext}
K_{yy} = -\dfrac{d-1}{r^{1-a}}\,e^t_y - {1\over 2r^a} \,e^r_y \,\gamma^{ij} \pt_t \gamma_{ij} - {r^a\over 2} \,e^t_y \,\gamma^{ij} \pt_r \gamma_{ij}\;.
\ee

Asymptotically, $\pt_t \gamma_{ij} \sim 1/r^{d+1-a/2}$ and $\pt_r \gamma_{ij} \sim 1/r^{d+1} $. For $a=2$, this is nothing but the requirement that the solution is asymptotically AdS, with the round metric on the conformal boundary. For asymptotically flat spacetimes, one of the defining properties is that all the derivatives of the metric perturbation decay with the same inverse power law of the radius. An asymptotic analysis of the $K=0$ condition reveals the large-$r$ scalings  $e^r_y \sim r^{a/2}$, $ e^t_y \sim 1/r^{d-1+a}$, so that the right hand side of \eqref{ext} is dominated by the first term:
\be
K_{yy} \approx -\dfrac{d-1}{r^{1-a}}\,e^t_y \;.
\ee   

Since $e_y \cdot \pt_t = - \,r^a\,e^t_y\,$, we learn that \eqref{bcon} can be satisfied provided the $C$-field is chosen with the boundary conditions
\be\label{boc}
C \rightarrow -\dfrac{1}{b}\,r(y) \,\pt_y\; \;\;\; {\rm as}\;\;\;\;y\rightarrow \infty\;. 
\ee
This is exactly the same result that was found for asymptotically AdS boundary conditions in \cite{prd}, justifying  the name `infall field' for the $C$-field. Similarly, the $M$-field satisfying \eqref{bcondm} will asymptotically factorize as in \eqref{factor} for the $C$-field given by \eqref{boc}. 

\bibliographystyle{style}
\bibliography{GPVC.bib}

\end{document}